\documentclass[prl,twocolumn,aps,showpacs,amsmath,amssymb]{revtex4-1}

\usepackage{graphicx}
\usepackage{epstopdf}
\usepackage{dcolumn}
\usepackage{bm}
\usepackage{amsmath}

\begin{document}


\title{Ultrafast terahertz probes of interacting dark excitons in chirality-specific semiconducting single-walled carbon nanotubes}

\author{Liang Luo, Ioannis Chatzakis${\dagger}$, Aaron Patz, Jigang Wang$^{*}$}

\affiliation{Department of Physics and Astronomy and Ames Laboratory-U.S. DOE, Iowa State University, Ames, Iowa 50011, USA.}

\date{\today}

\begin{abstract}
Ultrafast terahertz spectroscopy accesses the {\em dark} excitonic ground state in resonantly-excited (6,5) SWNTs via internal, direct dipole-allowed transitions between lowest lying dark-bright pair state $\sim$6 meV. An analytical model reproduces the response which enables quantitative analysis of transient densities of dark excitons and {\em e-h} plasma, oscillator strength, transition energy renormalization and dynamics. 
Non-equilibrium, yet stable, quasi-1D quantum states with dark excitonic correlations rapidly emerge even with increasing off-resonance photoexcitation and experience a unique crossover to complex phase-space filling of 
both dark and bright pair states, different from dense 2D/3D excitons influenced by the thermalization, cooling and ionization to free carriers.
\end{abstract}

\maketitle

Quasi-one-dimension (quasi-1D) excitons in single-walled carbon nanotubes (SWNTs), with large binding energies of 100s of meV, naturally arise from strong quantum confinement and reduced screening of electron-hole ({\em e-h}) pairs \cite{Feng,Maultzsch}. Their internal structure is characterized, in a 1D hydrogen atom-like description, by a center-of-mass momentum {\em K} and by internal quantum numbers (designated here as 1{\em s}, 2{\em s}, 2{\em p...}).  These strong excitonic behaviors manifest themselves in extensive static/ultrafast optical interband absorption/emission spectra in individually separated SWNTs, which are stable even in the presence of unbound {\em e-h} carriers from residual tube aggregation and/or metallic tubes \cite{YMa,JWang,Ostojic,Crochet}. However, unlike the hydrogen atom, the correlated {\em e-h} pairs in SWNTs evolve in a ``modified vacuum" with exotic, chiral symmetry arising from the underlying graphene lattice that gives two-fold degeneracy at  $K$ and $K'$ points. Coulomb interaction splits such ``doubling" into odd ({\em u}) and even ({\em g}) symmetry states entailing a series of bright and optically-forbidden, dark exciton pairs, including the lowest 1{\em s}({\em u}) and 1{\em s}({\em g}) \cite{Zhao, perebeinos} (Fig. 1(a)). 
Thus far, the dark ground state 1{\em s}({\em g}), hidden from both single- and two-photon optical interband transitions, is still largely unexplored.

While magneto-optics and light scattering measurements established the existence of dark excitons in SWNTs \cite{Srivastava,Matsunaga,Torrens}, the search for possible dipole-allowed, lowest transitions 1{\em s}({\em g})$\rightarrow$1{\em s}({\em u}) remains open. Particularly, as these linear probes do not depend critically on quasi-particle/exciton interaction and the associated fast dynamics, they provide little insight into interacting states that are characterized by the co-existence of both the dark $\&$ bright excitons and their complex interplay with {\em e-h} plasma. 
Exactly these aspects govern the radiative lifetime and photoluminesce (PL) efficiency, key for SWNT-based optoelectronic applications \cite{Xia,Zhao}. The lack of quantitative probes and scarce ultrafast measurements for dark states seriously limit their thorough understandings and perspectives of exploring related novel quantum phenomena, e.g., the excitonic Bose condensate \cite{Butov} and other exotic ground states.

Selective optical pump and THz probe technique represents a versatile spectroscopy tool that is extremely relevant for {\em quantitative} study of dark excitons. THz pulses directly couple the  1{\em s}({\em g})$\rightarrow$1{\em s}({\em u}) transitions [arrows, Fig. 1(a)], which represents a direct consequence and measure of excitonic pair correlations and resulting {\em dark} states in SWNTs. Being independent of momentum $K$, THz pulses measure genuine dark exciton population across entire $K$ space, while the available interband optical probes (PL, absorption), due to symmetry and momentum conservation, only detect a subset of excitons near $K$=0.
Additionally, the capability of resonant and off-resonant excitations by tuning the pump photon energy enables, among others, the study of excitons in single chirality tube and of exciton-plasma interaction.  
These critical aspects for dark exciton studies are absent in prior THz experiments of SWNTs. They reveal some other interesting low-energy electrodynamics, e.g., the absorption band centered $\sim$4 THz \cite{Kampfrath2}, and evidence for the internal transitions and non-Drude conductivity \cite{Xu2009}.     

In this Letter, we reveal lowest lying dark excitons and their interacting states in semiconducting SWNTs. With resonant excitation to the (6,5) $E_{22}$ interband transition at low temperature, transient THz spectra evidence strong photoinduced absorption centered $\sim$6 meV, whose transition energy, pump photon and temperature dependence, and dynamics manifest the observation of the 1{\em s}({\em g})$\rightarrow$1{\em s}({\em u}) transition.  
Most intriguingly, the 1{\em s}({\em g})$\rightarrow$1{\em s}({\em u}) oscillator signals emerge quasi-instantaneously even with increasing off-resonance photoexcitation and drop significantly above $\sim$130 $\mu$J cm$^{-2}$, at which the intra-excitonic resonances are {\em still stable with little shift or broadening}.  This appears to be distinctly different from dense 2D and 3D excitons that are influenced by the thermalization, cooling and ionization to free carriers \cite{Kaindl}. 
The robust non-equilibrium quasi-1D many-exciton states, instead, uniquely evolve from a predominant dark exciton population to complex phase-space filling of both dark and bright pair states.

\begin{figure}[tbp]
\begin{center}
\includegraphics[width=86mm]{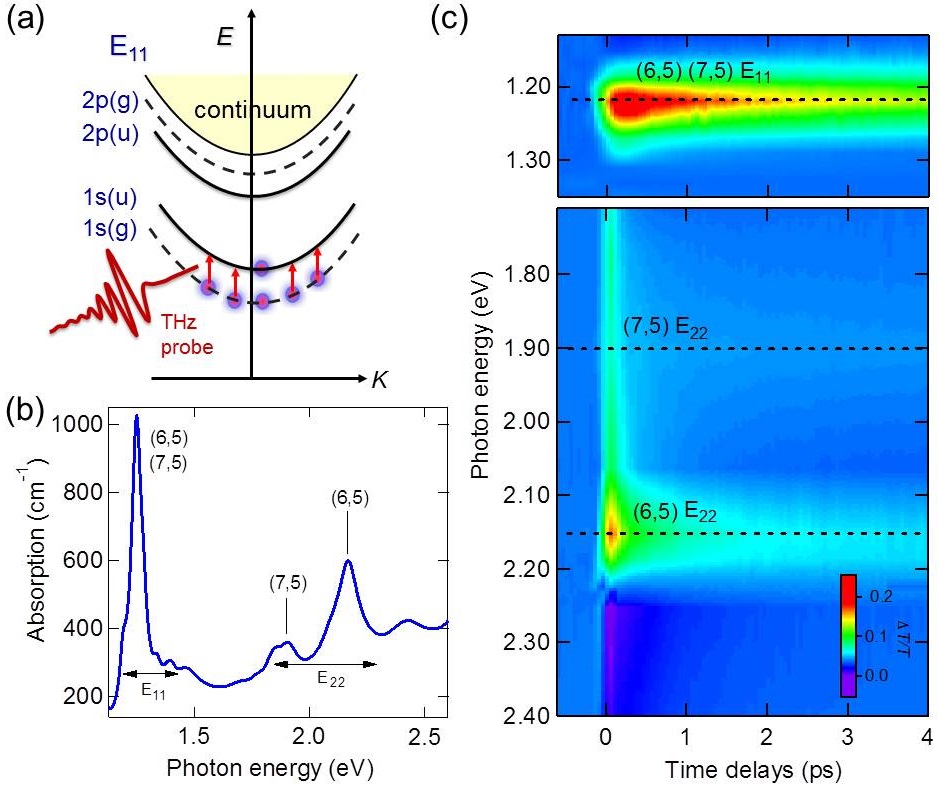}
\caption{(a) Schematic of two-particle {\em e-h} pair dispersion, illustrating the lowest lying 1{\em s}({\em g})$\rightarrow$1{\em s}({\em u}) intra-excitonic transitions (arrows) resonantly probed by THz pulses. Absorption spectra (b) and time/spectral-resolved differential transmission $\Delta T$/$T$ after 1.55 eV photoexcitation (c) of the SWNT sample at $T$=300 K. 
}
\label{Fig1}
\end{center}
\end{figure}

We study Co-Mo-catalyst grown SWNTs of mainly (6,5) and (7,5) chiralities embedded in a freestanding 50 $\mu$m Sodium Dodecylbenzenesulfonate (SDBS) film through drying a D$_{2}$O solution of SDBS-dispersed SWNTs (supplementary) \cite{Ogawa}. The absorption spectrum in Fig. 1(b) exhibits the distinct $E_{11}$ and $E_{22}$ quantized interband absorption peaks of dominant (6,5) and (7,5) chiralities, each associated with substantially prolonged relaxation times and significantly enhanced photobleaching, as shown in the differential transmission spectra (Fig. 1(c)). These observations are consistent with the extensive interband studies of high-quality chirality-enriched SWNT samples at visible and near-infrared spectral regions, which are explained by the joint effects of resonantly-enhanced radiative lifetime and exciton-exciton annihilation (EEA) \cite{Ostojic,YMa}.      

\begin{figure}[tbp]
\begin{center}
\includegraphics[width=86mm]{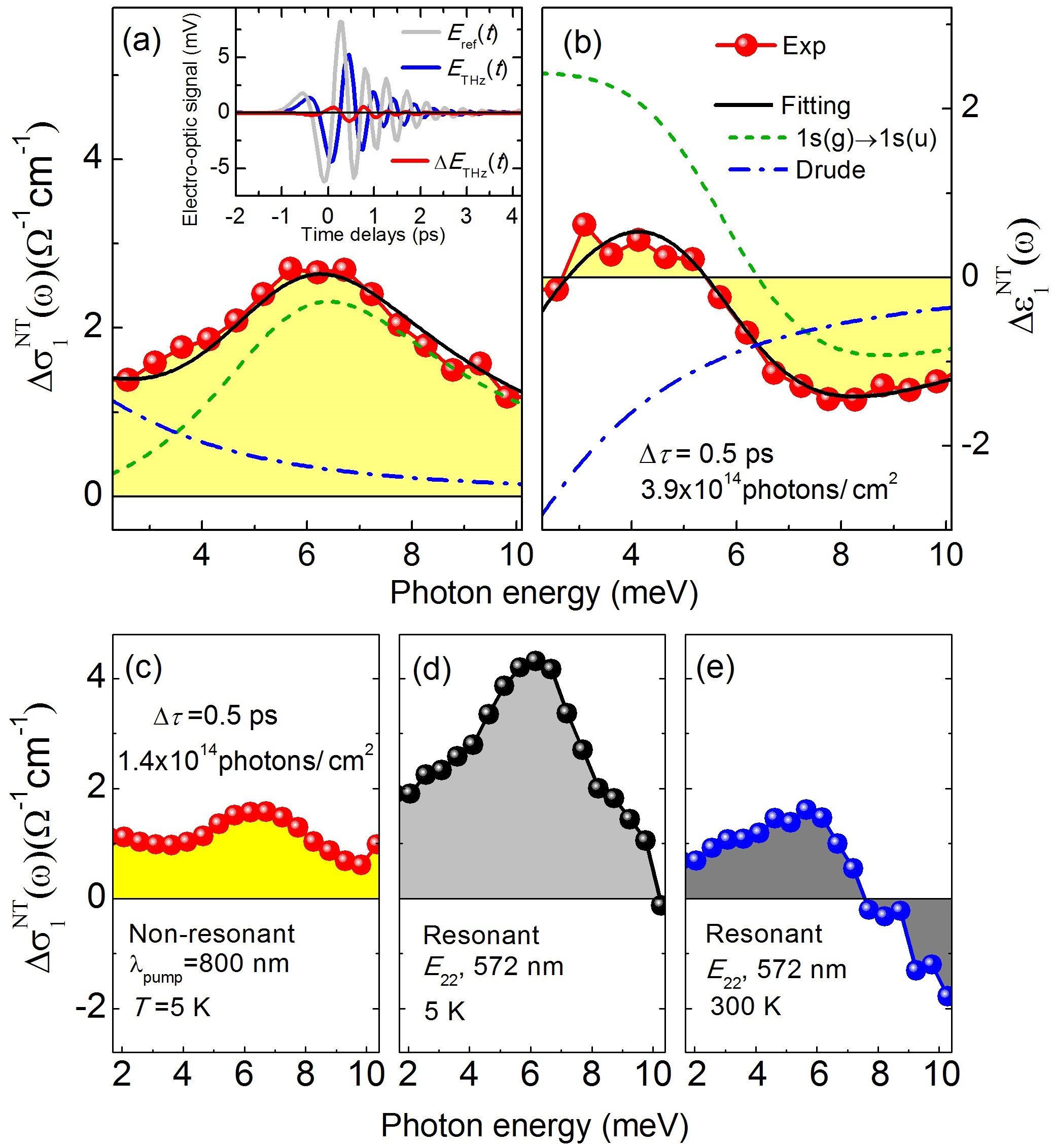}
\caption{Ultrafast THz spectra of $\Delta\sigma_{1}^{\text{NT}}(\omega)$ (a) and  $\Delta\varepsilon_{1}^{\text{NT}}(\omega)$ (b) (red dots), measured at pump-probe delay $\Delta\tau$=0.5 ps after 1.55 eV pumping. 
Thick line is the fit using the analytical model, Eq.($\ref{equ1}$), which is sum of the 1{\em s}({\em g})$\rightarrow$1{\em s}({\em u}) excitonic (dashed green) and unbound {\em e-h} responses (dash-dotted blue) (main text). Inset: THz fields transmitted (raw data). 
Comparison of photoinduced $\Delta\sigma_{1}^{\text{NT}}(\omega)$ at $\Delta\tau$=0.5 ps and under excitation density 1.4$\times10^{14}$ cm$^{-2}$: (c) Off-resonant pumping at 1.55 eV and $T$=5K; (d) on-resonant pumping at 2.17 eV and $T$=5K; (e) on-resonant pumping at 2.17 eV and $T$=300K.} 
\label{Fig2}
\end{center}
\end{figure}

Our optical pump, THz probe spectroscopy setup is driven by a 1 kHz Ti:Sapphire regenerative amplifier (40 fs, 800 nm center wavelength), which pumps an optical parametric amplifier and 1 mm thick $\langle 110\rangle$ ZnTe crystals (supplementary). 
The THz fields in time domain are measured, shown in the inset of Fig. 2(a), for transmission through reference (a clear aperture in our case) $E_\text{ref}(t)$ (gray), unexcited sample $E_\text{THz}(t)$ (blue), and its pump induced change $\Delta E_\text{THz}(t)$ after a pump-probe delay $\Delta \tau$=0.5 ps (red). Through the fast Fourier transformation and Fresnel equation, the full THz dielectric response is obtained and expressed as the real part of the conductivity, ${\sigma}_1^{\text{NT}}(\omega)$, and of the dielectric function, ${\varepsilon}_1^{\text{NT}}(\omega)$, which measures the absorbed power and the out-of-phase, inductive response, respectively \cite{Ioannis}. Effective medium theory is applied to obtain the SWNT dielectric function from the experimental data (supplementary).

Representative ultrafast THz responses ${\Delta\sigma}^{\text{NT}}_1(\omega)$ and ${\Delta\varepsilon_1^{\text{NT}}(\omega)}$ are shown in Figs. 2(a)-2(b) (red dots) at $\Delta\tau$=0.5 ps and $T$=5 K, after 1.55 eV pumping with photon density 3.9$\times10^{14}$ cm$^{-2}$. A strong photo-induced absorption appears in ${\Delta\sigma}_1^{\text{NT}}(\omega)$ within the time resolution with a broad, resonant spectral shape centered at $\sim$6 meV and a dispersive zero crossing occurs in ${\Delta\varepsilon_1^{\text{NT}}(\omega)}$ at slightly lower photon energy. These two features are characteristic of a well-defined driven THz oscillator while the shift between their spectral positions 
and the non-vanishing conductivity, albeit small, at the lowest probe energy $\sim$2 meV indicate additional low-frequency spectral weight, in the form of a coexisting unbound {\em e-h} plasma as discussed later. We emphasize three key facts to assign this resonance mainly from the 1{\em s}({\em g})$\rightarrow$1{\em s}({\em u}) transition of (6,5) tubes \cite{note2}. First, resonant photoexcitation of the (6,5) interband $E_{22}$ excitonic transitions at 2.17 eV, shown in Fig. 2(d), leads to significant enhancement of the transient THz resonance $\sim$6 meV, e.g., roughly three times as large as the 1.55 eV off-resonant excitation (Fig. 2(c)), for the same excitation photon density $n$=1.4$\times10^{14}$ cm$^{-2}$ per pulse. 
This clearly shows the excitonic origins of the THz resonance in (6,5) tubes. 
Second, raising the initial lattice temperature to $T$=300 K (Fig. 2(e)) diminishes the resonance as compared to $T$=5 K (Fig. 2(d)), for the same resonant excitation. Third, the resonance occurs in the transparent region of the unexcited sample 
and is close to the 1{\em s}({\em g})$\rightarrow$1{\em s}({\em u}) transition energy of (6,5) tubes indirectly extrapolated in high field magneto-optical studies \cite{Srivastava,Matsunaga}.  

For a quantitative analysis, a theoretical model is used to reproduce the experimentally determined complex THz dielectric functions. 
The model consists of two components: the THz dielectric function of the intra-excitonic transitions $\varepsilon^{1s}_{g\rightarrow u}(\omega)$ plus a Drude term of {\em e-h} plasma $\varepsilon_\mathit{eh}(\omega)$   
\begin{eqnarray}  \label{equ1}
\begin{aligned}
&\varepsilon^{\text{NT}}(\omega) = \varepsilon^{1s}_{g\rightarrow u}(\omega) + \varepsilon_\mathit{eh}(\omega)=\\
&[\varepsilon_{\infty}+ \frac{F^{1s}_{g\rightarrow u}}{(\omega^{1s}_{g\rightarrow u})^{2}- \omega^{2}-i\omega\Gamma}]-\frac {\omega_\text{p}^{2}} {\omega^{2}+i\omega\gamma}. 
\end{aligned}
\end{eqnarray}
For the first component (dashed green), $ F^{1s}_{g\rightarrow u}$ denotes effective transition strength of the intra-excitonic $\omega^{1s}_{g\rightarrow u}$ resonance as $F^{1s}_{g\rightarrow u}=f^{1s}_{g\rightarrow u}\cdot (\bigtriangleup \omega_\text{p}^{2})_{g\rightarrow u}^{1s}$, where $f$ is the oscillator strength and $\omega_\text{p}^{2}=ne^{2}/\varepsilon_{0}m$. The $\bigtriangleup \omega_\text{p}^{2}$ term, thereby, measures the population difference between the lowest dark and bright pair states (Fig. 1(a)), i.e., $\Delta n_\text{X}=n^{1s}_g-n^{1s}_u$. The second, Drude component (dash-dotted blue) is added with $\omega_{\text{p}}^{2}$ proportional to the density of unbound {\em e-h} density $n_\mathit{eh}$. $\varepsilon_{0}$ and $\varepsilon_{\infty}$ are the vacuum and background permittivity, and $e$ and $m$ are the electron charge and the effective exciton (or electron) mass obtained from \cite{Jorio}. Such composite THz response model (solid black lines) provides an excellent agreement with the low temperature data (red dots) in Figs. 2(a)-2(b) by varying the $\omega^{1s}_{g\rightarrow u}$ and $ F^{1s}_{g\rightarrow u}$, carrier density $n_\mathit{eh}$, and the exciton (carrier) broadening $\Gamma $ ($\gamma$). The best fit are strongly restrained by the requirement of simultaneously satisfying both dielectric responses ${\Delta\sigma}_1^{\text{NT}}(\omega)$ and ${\Delta\varepsilon_1^{\text{NT}}(\omega)}$, over a broad spectral range, and by the distinctly different spectral shapes of the excitonic oscillator and Drude carriers.  

\begin{figure}[tbp]
\begin{center}
\includegraphics[width=86mm]{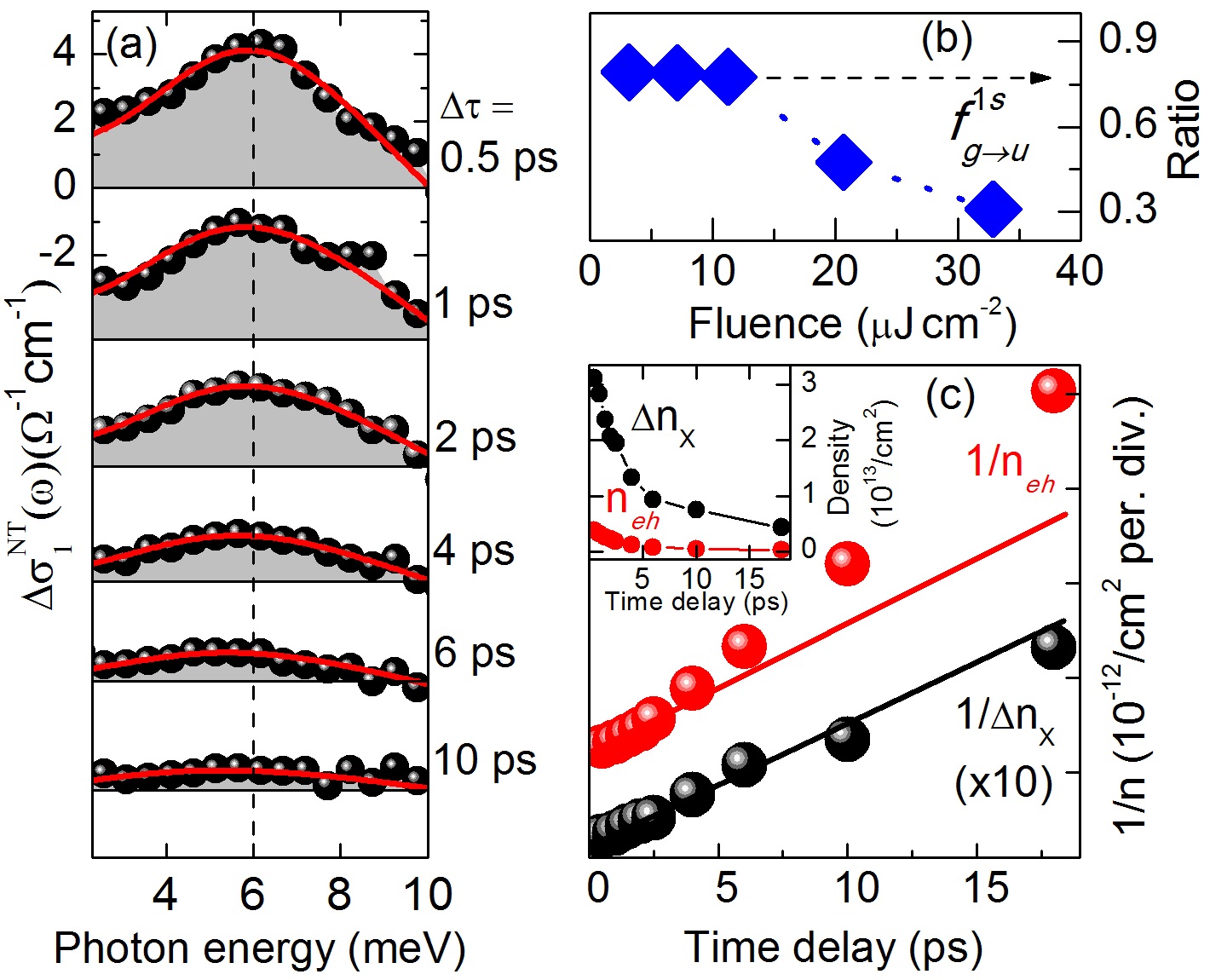}
\caption{(a) photo-induced conductivity changes $\Delta\sigma_{1}^{\text{NT}}(\omega)$ (black dots) at several pump-probe delays after resonant (6,5) $E_{22}$ excitation at 2.17 eV and $T$=5 K. Shown together are the model calcualtion (red lines) and peak positions (dash line). (b) The ratio $R=F^{1s}_{g\rightarrow u}$ $\cdot(\varepsilon_{0}m/e^{2})/(n_{\text{ph}}-n_{eh})$ as a function of fluence. (c) The reciprocal of exciton ($\Delta n_{\text{X}}$) and unbound {\em e-h} ($n_{eh}$) densities as a function of time delay together with fitting. The traces are vertically offset for clarity. Inset: temporal decay of $\Delta n_{\text{X}}$ and $n_{eh}$.}
\label{Fig3}
\end{center}
\end{figure}

Fig. 3(a) presents in detail of ultrafast THz conductivity spectra ${\Delta\sigma}_1^{\text{NT}} (\omega)$ for various time delays $\Delta \tau$ after the resonant $E_{22}$ photoexcitation at 2.17 eV. Pump fluence is 48 $\mu$J cm$^{-2}$ and $T$=5 K. The transient spectra are characterized by the $\omega^{1s}_{g\rightarrow u}$ resonance that retains its shape as the amplitude decays with time. The resonance peaks (dash line) exhibit no noticeable shift. The results can be fitted very well by the model described above (red lines). To further determine the exciton density $\Delta n_\text{X}=(F^{1s}_{g\rightarrow u}/f^{1s}_{g\rightarrow u}$)$\cdot (\varepsilon_{0}m/e^{2}$), the oscillator strength $f$ is needed for the 1{\em s}({\em g})$\rightarrow$1{\em s}({\em u}) transition which has never been determined before. This can be directly obtained here via fluence dependence of the ratio $R=F^{1s}_{g\rightarrow u}$ $\cdot(\varepsilon_{0}m/e^{2})/(n_{\text{ph}}-n_{eh})$ down to $I=$3$\mu$J cm$^{-2}$ in Fig. 3(b), where $n_\text{ph}$ is the actual absorbed photon density after taking into account the reflection and transmission of the pump beam in the optical path.  This ratio, with sufficiently weak photoexcitation, follows $R=f^{1s}_{g\rightarrow u}\cdot [1-\alpha(I)]/[1+\alpha(I)]$ with $\alpha=n^{1s}_u/n^{1s}_g$, considering strongly limited EEA in dilute exciton gas and fast inter-subband $E_{22}$-to-$E_{11}$ relaxation $\sim$40 fs \cite{Manzoni, Murakami}. The measurement clearly shows $R$ becomes mostly {\em pump-power-independent} below $\sim$10$\mu$J cm$^{-2}$ which indicates $\alpha\ll 1$ \cite{note4} and the ratio converges to a constant $f^{1s}_{g\rightarrow u} \approx$0.79. This is consistent with the fact that the optically generated exciton density is $\approx$1 per tube at $\sim$10$\mu$J cm$^{-2}$ estimated from the tube density and linear absorption. Increasing pump fluence to 10s of $\mu$J cm$^{-2}$ leads to a significant drop of the ratio (Fig. 3(b)), which can be attributed to efficient EEA that lowers the photon-exciton conversion efficiency. 

The obtained density $\Delta n_\text{X}$ and its relaxation dynamics (black dots) are well described by a bimolecular decay shown in Fig. 3(c). Although it exhibits a highly non-exponential profile over ps time scales (the inset), a reciprocal plot of 1/$\Delta n_\text{X}(t)$, instead, yields a simple straight line, i.e., 1/$\Delta n_\text{X}(t)\sim\beta t$ (black dots), which is the hallmark of the bimolecular decay given by $\frac{\mathrm{d} }{\mathrm{d} t}\Delta n_\text{X}(t)=-(1/2)\beta \Delta n_\text{X}^{2}(t)$ with decay rate of sheet density $\beta$=2.3$\times$10$^{-14}$ cm$^{2}$ ps$^{-1}$ \cite{YMa}.
The $1/n_\mathit{eh}$(t) decay (red dots), on the contrary, show large variation from the bimolecular behavior, which further underscores the excitonic origin of the $\Delta n_\text{X}$ response.
Furthermore, the strong similarity between the $\Delta n_\text{X}$(t) decay and firmly established exciton bimolecular kinetics in SWNTs also indicates $\Delta n_\text{X}$(t)$\approx n^{1s}_{g}$, i.e., it quantitatively follows the 1{\em s}({\em g}) exciton density. This shows that, under current pumping conditions, most excitons populate in the 1{\em s}({\em g}) ground state and $n^{1s}_{g} \gg n^{1s}_{u}$. Therefore, the resonant photoexcitation generates dark excitons $n^{1s}_{g}\sim$90\% of the total density, much larger than $n_\mathit{eh}$ (inset, Fig. 3(c)).

\begin{figure}[tbp]
\begin{center}
\includegraphics[width=86mm]{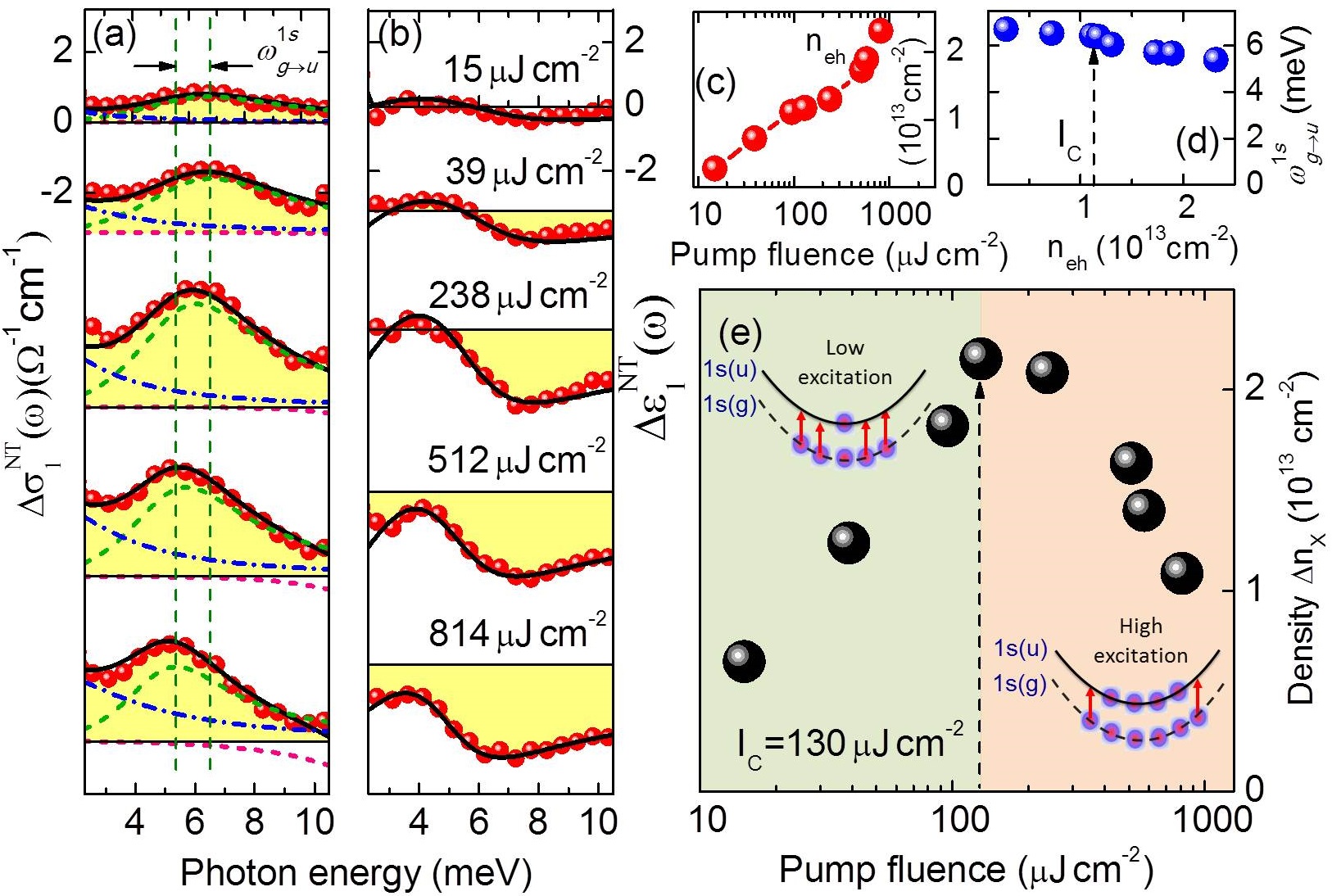}
\caption{(a)-(b) photo-induced $\Delta\sigma_{1}^{\text{NT}}(\omega)$ and $\Delta\varepsilon_{1}^{\text{NT}}(\omega)$ changes (red dots) at various pump fluences after 1.55 eV excitation. $\Delta \tau$=0.5 ps and $T$=5K. The model fitting (black lines) is divided (see text) and the dashed lines indicate the shift of the 1{\em s}({\em g})$\rightarrow$1{\em s}({\em u}) peak positions. (c) The unbound {\em e-h} density as a function of pump fluence. (d) The resonance peak as a function of unbound {\em e-h} density. (e)  Fluence dependence of the exciton signal $\Delta n_\text{X}$, which is divided into two regimes at $I_C \sim$130 $\mu$J cm$^{-2}$ (dash arrow). }
\label{Fig4}
\end{center}
\end{figure}

Increasing off-resonant excitation reveals the complex many-particle interacting state. 
Figs. 4(a)-(b) shows $\Delta\sigma_{1}^{\text{NT}}(\omega)$ and $\Delta\varepsilon_{1}^{\text{NT}}(\omega)$ for various pump fluences after 1.55 eV photoexcitation at $T$=5 K. The spectra at $\Delta \tau$=0.5 ps clearly exhibit the $\omega^{1s}_{g\rightarrow u}$ resonance, despite 100s of meV detuning away from the excitonic resonances, which dominates {\em e-h} plasma especially at low pumping. This reveals an ultrafast {\em sub-ps} formation of non-equilibrium 1D dark excitonic correlations, much faster than the 100s of ps thermalization of bright excitons to the dark ground states seen in time-resolved PL \cite{Matsunaga2}. The difference occurs since the initial electronically correlated states are mostly hidden from those interband probes. The formation also appears to be much more robust than 2D/3D excitons that are strongly influenced by the co-existing {\em e-h} plasma with 100s of ps cooling time under the off-resonant pumping \cite{Kaindl}. The composite THz model (black lines) again consistently reproduces very well the experimental results, which are divided into individual components in the same manner as Fig. 2(a): the intraexcitonic 1{\em s}({\em g})$\rightarrow$1{\em s}({\em u}) (green lines) and unbound {\em e-h} carriers (blue lines).  Note the bleaching component of the 4 THz band (pink lines) \cite{Kampfrath2,note3} is added and {\em only} contributes at high pumping above 512 $\mu$J cm$^{-2}$ (supplementary). Two salient features are visible with increasing pump fluence: \\
(1)	The first feature is that the $\omega^{1s}_{g\rightarrow u}$ resonance in $\Delta\sigma_{1}^{\text{NT}}(\omega)$ shifts to lower frequency (Fig. 4(d)), illustrated by two dash lines in Fig. 4(a), $\sim$20 \% change from 15 to 814 $\mu$J cm$^{-2}$.  We attribute this to the critical role of plasma-exciton interaction in the renormalization of the excitonic levels, considering the unbound {\em e-h} density significantly increases in the off-resonance pumping, e.g., $ n_\mathit{eh}$ can reach $\sim$2.5$\times 10^{13}$cm$^{-2}$, comparable to exciton density, shown in Figs. 4(c) and 4(e). Note that the $\omega^{1s}_{g\rightarrow u}$ shift is not seen in the system of a predominant exciton population, shown in Fig. 3 for the resonant pumping, which has larger exciton density but one order of magnitude smaller $ n_\mathit{eh}$.  
This reconciles some contradictory results regarding the exciton stability in the literature, e.g., photoluminescence spectra of dense excitons have shown no excitonic shift up to complete absorption saturation \cite{Ostojic,Murakami}, while ultra-violet pumping results in some effects \cite{Crochet}. This can be underpinned unambiguously here to the {\em e-h} plasma-exciton interaction. \\
(2) The second feature is that the exciton density $\Delta n_\text{X}$ exhibits a distinct non-monotonic variation with increasing photoexcitation (Fig. 4(e)), which peaks at $I_C \sim$130 $\mu$J cm$^{-2}$ followed by a significant decrease with further increasing pump fluence. However, most interestingly, the intra-excitonic resonance $\omega^{1s}_{g\rightarrow u}$ exhibits {\em little shift or broadening} at $I_C$ (\textless 5$\%$, dash arrow in Fig. 4(d)), indicating that there is {\em no obvious exciton ionization to unbound e-h plasma}. 
This appears to be fundamentally different from dense 2D and 3D excitons where the strong decrease of excitonic signals occurs {\em only} upon the ionization of excitons manifested by a significant shift, broadening, and ultimately the disappearance of the THz resonances. The quasi-1D many-body state in SWNTs reveal, instead, the evolution from a predominant dark exciton population in the 1{\em s}({\em g}) to phase space filling of both the lowest dark and bright exciton pair states \cite{note}. 
Such crossover at $I_C$ is illustrated by two shaded colors in Fig. 4(e): below $I_C$, photoexcited excitons primarily populate in the lowest lying dark state 1{\em s}({\em g}) that leads to a rapid rise of $\Delta n_\text{X}$ with increasing pump fluence; above $I_C$, photoexcited excitons have larger probability to populate the 1{\em s}({\em u}) state rather than the 1{\em s}({\em g}) ground state, due to electronic heating of the many-body systems, which is responsible for the reduction of the $\omega^{1s}_{g\rightarrow u}$ resonance via Pauli blocking of the transition. The 1{\em s}({\em g})/1{\em s}({\em u}) exciton pairs, being well-isolated from higher lying exciton levels and continuum ($>$200 meV) \cite{JWang,Feng,Maultzsch}, is stable against high density ionization. 

In conclusion, we provide the first insights into the chirality-specific THz response of non-equilibrium excitonic correlations and dynamics from the dark ground states in SWNTs. The THz 1{\em s}({\em g})$\rightarrow$1{\em s}({\em u}) resonant probes revealed, being independent of ground state symmetry and momentum conservation restrictions, identify that strong electronic correlation regulates the sub-ps formation of 1D dark excitons beyond the previously-held slow thermalization/cooling scenario. This may also evolve into a benchmark approach for quantitative exciton management in SWNT-based device development, and motivates for fundamental quantum phase discovery of excitons \cite{Butov} and other strongly correlated excitations \cite{li-2013}.       

This work was supported by the US Department of Energy, Office of Basic Energy Science, Division of Materials Sciences and Engineering (Ames Laboratory is operated for the US Department of Energy by Iowa State University under Contract No. DE-AC02-07CH11358).

\noindent$^{*}$Corresponding author.
\\jgwang@iastate.edu; jwang@ameslab.gov

\noindent$^{\dagger}$Present address: Ming Hsieh Department of Electrical Engineering, University of Southern California, 3737 Watt Way, Los Angeles, CA 90089-0271

\bibliographystyle{apsrev}

\begin{thebibliography}{99}
\bibitem{note2} A phonon resonance scenario can be excluded since the resonances (or periodical oscillations) are absent in static THz conductivity spectra (or the photoinduced temporal traces).
\bibitem{note4} The small $n^{1s}_u/n^{1s}_g$ is consistent with the estimation from time-averaged photoluminance $\approx$0.06$-$0.08 below 10K \cite{Matsunaga2}.
\bibitem{note} Note that the photoexcited exciton density is still below the expected Mott density in SWNTs \cite{Murakami}.
\end{thebibliography}

\newpage

\end{document}